\documentclass[longauth]{aa} 

\usepackage{graphicx}
\usepackage{microtype}
\usepackage{placeins}
\usepackage{txfonts}
\usepackage{xspace}
\usepackage{color}
\usepackage{url}
\usepackage{hyperref}
\usepackage[flushleft]{threeparttable}
\newcommand {\bc}{\begin {center}}
\newcommand {\ec}{\end {center}}
\newcommand {\be}{\begin {equation}}
\newcommand {\ee}{\end {equation}}
\newcommand {\beq}{\begin {eqnarray}}
\newcommand {\eeq}{\end {eqnarray}}

\newcommand {\nustar}{\textit{NuSTAR}\xspace}
\newcommand {\nicer}{{NICER}\xspace}
\newcommand {\ixpe}{\textit{IXPE}\xspace}

\newcommand {\rx}{\mbox{RX~J0440.9+4431}\xspace}
\newcommand {\lsv}{\mbox{LS~V~+44~17}\xspace}

\begin{document} 

\title{Complex variations in X-ray polarization in the X-ray pulsar \mbox{LS~V~+44~17}/\rx}

\titlerunning{X-ray polarimetry of the X-ray pulsar \mbox{LS~V~+44~17}}
\authorrunning{V.~Doroshenko et al.}

\author{
Victor Doroshenko \inst{\ref{in:Tub}}
\and Juri Poutanen \inst{\ref{in:UTU}}
\and Jeremy~Heyl \inst{\ref{in:UBC}}
\and Sergey~S.~Tsygankov \inst{\ref{in:UTU}} 
\and Ilaria~Caiazzo \inst{\ref{in:Caltech_dept}} 
\and Roberto~Turolla \inst{\ref{in:UniPD},\ref{in:MSSL}}  
\and Alexandra~Veledina \inst{\ref{in:UTU},\ref{in:KHT}} 
\and Martin~C.~Weisskopf \inst{\ref{in:NASA-MSFC}} 
\and Sofia~V.~Forsblom \inst{\ref{in:UTU}}
\and Denis Gonz{\'a}lez-Caniulef \inst{\ref{in:IRAP}}
\and Vladislav~Loktev \inst{\ref{in:UTU}}
\and Christian~Malacaria \inst{\ref{in:ISSI}}
\and Alexander~A.~Mushtukov \inst{\ref{in:Oxford}}
\and Valery~F.~Suleimanov \inst{\ref{in:Tub}}  
\and Alexander~A.~Lutovinov \inst{\ref{in:IKI}}
\and Ilya~A.~Mereminskiy \inst{\ref{in:IKI}}
\and Sergey~V.~Molkov\inst{\ref{in:IKI}} 
\and Alexander~Salganik \inst{\ref{in:SPBU},\ref{in:IKI}} 
\and Andrea~Santangelo \inst{\ref{in:Tub}} 
\and Andrei~V.~Berdyugin \inst{\ref{in:UTU}}
\and Vadim~Kravtsov \inst{\ref{in:UTU}}
\and Anagha P. Nitindala \inst{\ref{in:UTU}} 
\and Iv\'an~Agudo \inst{\ref{in:CSIC-IAA}}
\and Lucio~A.~Antonelli \inst{\ref{in:INAF-OAR},\ref{in:ASI-SSDC}} 
\and Matteo~Bachetti \inst{\ref{in:INAF-OAC}} 
\and Luca~Baldini  \inst{\ref{in:INFN-PI},      \ref{in:UniPI}} 
\and Wayne~H.~Baumgartner  \inst{\ref{in:NASA-MSFC}} 
\and Ronaldo~Bellazzini  \inst{\ref{in:INFN-PI}} 
\and Stefano~Bianchi \inst{\ref{in:UniRoma3}}  
\and Stephen~D.~Bongiorno \inst{\ref{in:NASA-MSFC}} 
\and Raffaella~Bonino  \inst{\ref{in:INFN-TO},\ref{in:UniTO}}
\and Alessandro~Brez  \inst{\ref{in:INFN-PI}} 
\and Niccol\`{o}~Bucciantini 
\inst{\ref{in:INAF-Arcetri},\ref{in:UniFI},\ref{in:INFN-FI}} 
\and Fiamma~Capitanio \inst{\ref{in:INAF-IAPS}}
\and Simone~Castellano \inst{\ref{in:INFN-PI}}  
\and Elisabetta~Cavazzuti \inst{\ref{in:ASI}} 
\and Chien-Ting~Chen \inst{\ref{in:USRA-MSFC}}
\and Stefano~Ciprini \inst{\ref{in:INFN-Roma2},\ref{in:ASI-SSDC}}
\and Enrico~Costa \inst{\ref{in:INAF-IAPS}} 
\and Alessandra~De~Rosa \inst{\ref{in:INAF-IAPS}} 
\and Ettore~Del~Monte \inst{\ref{in:INAF-IAPS}} 
\and Laura~Di~Gesu \inst{\ref{in:ASI}} 
\and Niccol\`{o}~Di~Lalla \inst{\ref{in:Stanford}}
\and Alessandro~Di~Marco \inst{\ref{in:INAF-IAPS}}
\and Immacolata~Donnarumma \inst{\ref{in:ASI}}
\and Michal~Dov\v{c}iak \inst{\ref{in:CAS-ASU}}
\and Steven~R.~Ehlert \inst{\ref{in:NASA-MSFC}}  
\and Teruaki~Enoto \inst{\ref{in:RIKEN}}
\and Yuri~Evangelista \inst{\ref{in:INAF-IAPS}}
\and Sergio~Fabiani \inst{\ref{in:INAF-IAPS}}
\and Riccardo~Ferrazzoli \inst{\ref{in:INAF-IAPS}} 
\and Javier~A.~Garc{\'i}a \inst{\ref{in:Caltech}}
\and Shuichi~Gunji\inst{\ref{in:Yamagata}} 
\and Kiyoshi~Hayashida \inst{\ref{in:Osaka}}\thanks{Deceased} 
\and Wataru~Iwakiri \inst{\ref{in:Chiba}} 
\and Svetlana~G.~Jorstad \inst{\ref{in:BU},\ref{in:SPBU}} 
\and Philip~Kaaret \inst{\ref{in:NASA-MSFC}}  
\and Vladimir~Karas \inst{\ref{in:CAS-ASU}}
\and Fabian~Kislat \inst{\ref{in:UNH}} 
\and Takao~Kitaguchi  \inst{\ref{in:RIKEN}} 
\and Jeffery~J.~Kolodziejczak \inst{\ref{in:NASA-MSFC}} 
\and Henric~Krawczynski  \inst{\ref{in:WUStL}}
\and Fabio~La~Monaca \inst{\ref{in:INAF-IAPS}} 
\and Luca~Latronico  \inst{\ref{in:INFN-TO}} 
\and Ioannis~Liodakis \inst{\ref{in:FINCA}}
\and Simone~Maldera \inst{\ref{in:INFN-TO}}  
\and Alberto~Manfreda \inst{\ref{INFN-NA}}
\and Fr\'{e}d\'{e}ric~Marin \inst{\ref{in:Strasbourg}} 
\and Andrea~Marinucci \inst{\ref{in:ASI}} 
\and Alan~P.~Marscher \inst{\ref{in:BU}} 
\and Herman~L.~Marshall \inst{\ref{in:MIT}}
\and Francesco~Massaro \inst{\ref{in:INFN-TO},\ref{in:UniTO}} 
\and Giorgio~Matt  \inst{\ref{in:UniRoma3}}  
\and Ikuyuki~Mitsuishi \inst{\ref{in:Nagoya}} 
\and Tsunefumi~Mizuno \inst{\ref{in:Hiroshima}} 
\and Fabio~Muleri \inst{\ref{in:INAF-IAPS}} 
\and Michela~Negro \inst{\ref{in:UMBC},\ref{in:NASA-GSFC},\ref{in:CRESST}} 
\and Chi-Yung~Ng \inst{\ref{in:HKU}}
\and Stephen~L.~O'Dell \inst{\ref{in:NASA-MSFC}}  
\and Nicola~Omodei \inst{\ref{in:Stanford}}
\and Chiara~Oppedisano \inst{\ref{in:INFN-TO}}  
\and Alessandro~Papitto \inst{\ref{in:INAF-OAR}}
\and George~G.~Pavlov \inst{\ref{in:PSU}}
\and Abel~L.~Peirson \inst{\ref{in:Stanford}}
\and Matteo~Perri \inst{\ref{in:ASI-SSDC},\ref{in:INAF-OAR}}
\and Melissa~Pesce-Rollins \inst{\ref{in:INFN-PI}} 
\and Pierre-Olivier~Petrucci \inst{\ref{in:Grenoble}} 
\and Maura~Pilia \inst{\ref{in:INAF-OAC}} 
\and Andrea~Possenti \inst{\ref{in:INAF-OAC}} 
\and Simonetta~Puccetti \inst{\ref{in:ASI-SSDC}}
\and Brian~D.~Ramsey \inst{\ref{in:NASA-MSFC}}  
\and John~Rankin \inst{\ref{in:INAF-IAPS}} 
\and Ajay~Ratheesh \inst{\ref{in:INAF-IAPS}} 
\and Oliver~J.~Roberts \inst{\ref{in:USRA-MSFC}}
\and Roger~W.~Romani \inst{\ref{in:Stanford}}
\and Carmelo~Sgr\`{o} \inst{\ref{in:INFN-PI}}  
\and Patrick~Slane \inst{\ref{in:CfA}}  
\and Paolo~Soffitta \inst{\ref{in:INAF-IAPS}} 
\and Gloria~Spandre \inst{\ref{in:INFN-PI}} 
\and Douglas~A.~Swartz \inst{\ref{in:USRA-MSFC}}
\and Toru~Tamagawa \inst{\ref{in:RIKEN}}
\and Fabrizio~Tavecchio \inst{\ref{in:INAF-OAB}}
\and Roberto~Taverna \inst{\ref{in:UniPD}} 
\and Yuzuru~Tawara \inst{\ref{in:Nagoya}}
\and Allyn~F.~Tennant \inst{\ref{in:NASA-MSFC}}  
\and Nicholas~E.~Thomas \inst{\ref{in:NASA-MSFC}}  
\and Francesco~Tombesi  \inst{\ref{in:UniRoma2},\ref{in:INFN-Roma2},\ref{in:UMd}}
\and Alessio~Trois \inst{\ref{in:INAF-OAC}}
\and Jacco~Vink \inst{\ref{in:Amsterdam}}
\and Kinwah~Wu \inst{\ref{in:MSSL}}
\and Fei~Xie \inst{\ref{in:GSU},\ref{in:INAF-IAPS}}
\and Silvia~Zane  \inst{\ref{in:MSSL}}
          }
          
\institute{Institut f\"ur Astronomie und Astrophysik, Universit\"at T\"ubingen, Sand 1, D-72076 T\"ubingen, Germany \label{in:Tub}
\email{doroshv@astro.uni-tuebingen.de}
\and Department of Physics and Astronomy, FI-20014 University of Turku,  Finland \label{in:UTU}
\and 
University of British Columbia, Vancouver, BC V6T 1Z4, Canada \label{in:UBC}
\and Division of Physics, Mathematics and Astronomy, California Institute of Technology, 
Pasadena, CA 91125, USA \label{in:Caltech_dept} 
\and 
Dipartimento di Fisica e Astronomia, Universit\`{a} degli Studi di Padova, Via Marzolo 8, 35131 Padova, Italy \label{in:UniPD}
\and 
Mullard Space Science Laboratory, University College London, Holmbury St Mary, Dorking, Surrey RH5 6NT, UK \label{in:MSSL}
\and
Nordita, KTH Royal Institute of Technology and Stockholm University, Hannes Alfv\'{e}ns v\"{a}g 12, SE-106\,91 Stockholm, Sweden\label{in:KHT}
\and 
NASA Marshall Space Flight Center, Huntsville, AL 35812, USA \label{in:NASA-MSFC}
\and Institut de Recherche en Astrophysique et Plan\'etologie, UPS-OMP, CNRS, CNES, 9 avenue du Colonel Roche, BP 44346 31028, Toulouse CEDEX 4, France 
\label{in:IRAP} 
\and 
International Space Science Institute, Hallerstrasse 6, 3012 Bern, Switzerland \label{in:ISSI}
\and 
Astrophysics, Department of Physics, University of Oxford, Denys Wilkinson Building, Keble Road, Oxford OX1 3RH, UK \label{in:Oxford}
\and Space Research Institute (IKI) of the Russian Academy of Sciences, Profsoyuznaya Str. 84/32, Moscow 117997, Russia \label{in:IKI}
\and 
Department of Astrophysics, St. Petersburg State University, Universitetsky pr. 28, Petrodvoretz, 198504 St. Petersburg, Russia \label{in:SPBU} 
\and 
Instituto de Astrof\'{i}sicade Andaluc\'{i}a -- CSIC, Glorieta de la Astronom\'{i}a s/n, 18008 Granada, Spain \label{in:CSIC-IAA}
\and 
INAF Osservatorio Astronomico di Roma, Via Frascati 33, 00040 Monte Porzio Catone (RM), Italy \label{in:INAF-OAR}  
\and 
Space Science Data Center, Agenzia Spaziale Italiana, Via del Politecnico snc, 00133 Roma, Italy \label{in:ASI-SSDC}
 \and
INAF Osservatorio Astronomico di Cagliari, Via della Scienza 5, 09047 Selargius (CA), Italy  \label{in:INAF-OAC}
\and 
Istituto Nazionale di Fisica Nucleare, Sezione di Pisa, Largo B. Pontecorvo 3, 56127 Pisa, Italy \label{in:INFN-PI}
\and  
Dipartimento di Fisica, Universit\`{a} di Pisa, Largo B. Pontecorvo 3, 56127 Pisa, Italy \label{in:UniPI}
\and 
Dipartimento di Matematica e Fisica, Universit\`a degli Studi Roma Tre, via della Vasca Navale 84, 00146 Roma, Italy  \label{in:UniRoma3}
\and  
Istituto Nazionale di Fisica Nucleare, Sezione di Torino, Via Pietro Giuria 1, 10125 Torino, Italy  \label{in:INFN-TO}      
\and  
Dipartimento di Fisica, Universit\`{a} degli Studi di Torino, Via Pietro Giuria 1, 10125 Torino, Italy \label{in:UniTO} 
\and   
INAF Osservatorio Astrofisico di Arcetri, Largo Enrico Fermi 5, 50125 Firenze, Italy 
\label{in:INAF-Arcetri} 
\and  
Dipartimento di Fisica e Astronomia, Universit\`{a} degli Studi di Firenze, Via Sansone 1, 50019 Sesto Fiorentino (FI), Italy \label{in:UniFI} 
\and   
Istituto Nazionale di Fisica Nucleare, Sezione di Firenze, Via Sansone 1, 50019 Sesto Fiorentino (FI), Italy \label{in:INFN-FI}
\and 
Agenzia Spaziale Italiana, Via del Politecnico snc, 00133 Roma, Italy \label{in:ASI}
\and 
Science and Technology Institute, Universities Space Research Association, Huntsville, AL 35805, USA \label{in:USRA-MSFC}
\and 
Istituto Nazionale di Fisica Nucleare, Sezione di Roma ``Tor Vergata'', Via della Ricerca Scientifica 1, 00133 Roma, Italy 
 \label{in:INFN-Roma2}
 \and 
INAF Istituto di Astrofisica e Planetologia Spaziali, Via del Fosso del Cavaliere 100, 00133 Roma, Italy \label{in:INAF-IAPS}
\and 
Department of Physics and Kavli Institute for Particle Astrophysics and Cosmology, Stanford University, Stanford, California 94305, USA  \label{in:Stanford}
\and 
Astronomical Institute of the Czech Academy of Sciences, Bo\v{c}n\'{i} II 1401/1, 14100 Praha 4, Czech Republic \label{in:CAS-ASU}
\and 
RIKEN Cluster for Pioneering Research, 2-1 Hirosawa, Wako, Saitama 351-0198, Japan \label{in:RIKEN}
\and 
California Institute of Technology, Pasadena, CA 91125, USA \label{in:Caltech}
\and 
Yamagata University,1-4-12 Kojirakawa-machi, Yamagata-shi 990-8560, Japan \label{in:Yamagata}
\and 
Osaka University, 1-1 Yamadaoka, Suita, Osaka 565-0871, Japan \label{in:Osaka}
\and 
International Center for Hadron Astrophysics, Chiba University, Chiba 263-8522, Japan \label{in:Chiba}
\and
Institute for Astrophysical Research, Boston University, 725 Commonwealth Avenue, Boston, MA 02215, USA \label{in:BU} 
\and 
Department of Physics and Astronomy and Space Science Center, University of New Hampshire, Durham, NH 03824, USA \label{in:UNH} 
\and 
Physics Department and McDonnell Center for the Space Sciences, Washington University in St. Louis, St. Louis, MO 63130, USA \label{in:WUStL}
\and 
Finnish Centre for Astronomy with ESO,  20014 University of Turku, Finland \label{in:FINCA}
\and 
Istituto Nazionale di Fisica Nucleare, Sezione di Napoli, Strada Comunale Cinthia, 80126 Napoli, Italy \label{INFN-NA}
\and 
Universit\'{e} de Strasbourg, CNRS, Observatoire Astronomique de Strasbourg, UMR 7550, 67000 Strasbourg, France \label{in:Strasbourg}
\and 
MIT Kavli Institute for Astrophysics and Space Research, Massachusetts Institute of Technology, 77 Massachusetts Avenue, Cambridge, MA 02139, USA \label{in:MIT}
\and 
Graduate School of Science, Division of Particle and Astrophysical Science, Nagoya University, Furo-cho, Chikusa-ku, Nagoya, Aichi 464-8602, Japan \label{in:Nagoya}
\and 
Hiroshima Astrophysical Science Center, Hiroshima University, 1-3-1 Kagamiyama, Higashi-Hiroshima, Hiroshima 739-8526, Japan \label{in:Hiroshima}
\and  University of Maryland, Baltimore County, Baltimore, MD 21250, USA \label{in:UMBC}
\and NASA Goddard Space Flight Center, Greenbelt, MD 20771, USA  \label{in:NASA-GSFC}
\and Center for Research and Exploration in Space Science and Technology, NASA/GSFC, Greenbelt, MD 20771, USA  \label{in:CRESST}
\and 
Department of Physics, University of Hong Kong, Pokfulam, Hong Kong \label{in:HKU}
\and 
Department of Astronomy and Astrophysics, Pennsylvania State University, University Park, PA 16801, USA \label{in:PSU}
\and 
Universit\'{e} Grenoble Alpes, CNRS, IPAG, 38000 Grenoble, France \label{in:Grenoble}
\and 
Center for Astrophysics, Harvard \& Smithsonian, 60 Garden St, Cambridge, MA 02138, USA \label{in:CfA} 
\and 
INAF Osservatorio Astronomico di Brera, via E. Bianchi 46, 23807 Merate (LC), Italy \label{in:INAF-OAB}
\and
Dipartimento di Fisica, Universit\`{a} degli Studi di Roma ``Tor Vergata'', Via della Ricerca Scientifica 1, 00133 Roma, Italy \label{in:UniRoma2}
\and
Department of Astronomy, University of Maryland, College Park, Maryland 20742, USA \label{in:UMd}
\and 
Anton Pannekoek Institute for Astronomy \& GRAPPA, University of Amsterdam, Science Park 904, 1098 XH Amsterdam, The Netherlands  \label{in:Amsterdam}
\and 
Guangxi Key Laboratory for Relativistic Astrophysics, School of Physical Science and Technology, Guangxi University, Nanning 530004, China \label{in:GSU}
}

\abstract{
We report on Imaging X-ray polarimetry explorer (\ixpe) observations of the Be-transient X-ray pulsar {LS~V~+44~17}/\rx made at two luminosity levels during the giant outburst in January--February 2023. 
Considering the observed spectral variability and changes in the pulse profiles, the source was likely caught in supercritical and subcritical states with significantly different emission-region geometry, associated with the presence of accretion columns and hot spots, respectively. 
We focus here on the pulse-phase-resolved polarimetric analysis and find that the observed dependencies of the polarization degree and polarization angle (PA) on the pulse phase are indeed drastically different for the two observations. 
The observed differences, if interpreted within the framework of the rotating vector model (RVM), imply dramatic variations in the spin axis inclination, the position angle, and the magnetic colatitude by tens of degrees within the space of just a few days. 
We suggest that the apparent changes in the observed PA phase dependence are predominantly related to the presence of an unpulsed polarized component in addition to the polarized radiation associated with the pulsar itself. 
We then show that the observed PA phase dependence in both observations can  be explained with a single set of RVM parameters defining the pulsar's geometry. 
We also suggest that the additional polarized component is likely produced by scattering of the pulsar radiation in the equatorial disk wind. }


\maketitle
%
%

\section{Introduction}

The Be/X-ray binary (BeXRB) {LS V +44 17}/\rx was  discovered and identified as a candidate X-ray binary at $\sim$3.2\,kpc in the \textit{ROSAT} survey \citep{1997A&A...323..853M}.
The discovery of hard X-ray emission and pulsations with a period of about 206\,s 
confirmed the source is an X-ray pulsar (XRP,  \citealt{1999MNRAS.306..100R}).
The properties of the optical counterpart were investigated in detail by  \citet{2005A&A...440.1079R}, who classified it as a Be star of class B0.2V and reported on brightness and H$\alpha$ line profile variability typical for this class of sources. 
\citet{2005A&A...440.1079R} also estimated the distance to the source as being around 3.3\,kpc; this has  recently been revised to $2.4\pm0.1$\,kpc based on \textit{Gaia}~DR3 data \citep{2021AJ....161..147B}.

\begin{figure}
\centering
\includegraphics[width=0.9\columnwidth]{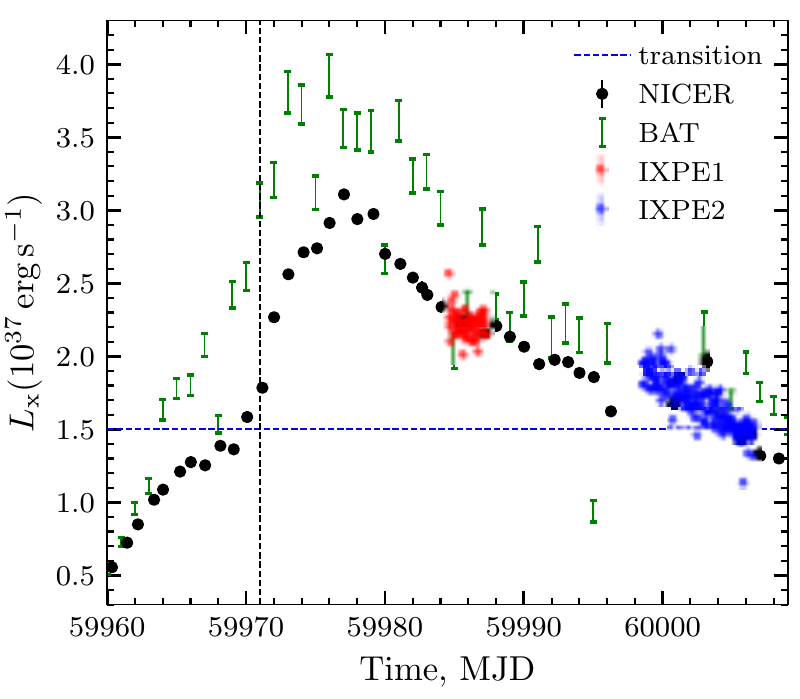}
\caption{Light curve of the 2023 outburst of \rx as observed by the facilities indicated in the legend and described in the main text. 
The vertical dashed line marks the probable date of the transition to a supercritical regime (around MJD~59971, i.e., January 27, 2023), while the horizontal dashed line marks the approximate luminosity level at the transition. 
}
\label{fig:lc}
\end{figure}

\begin{figure}
\centering
\includegraphics[width=0.9\columnwidth]{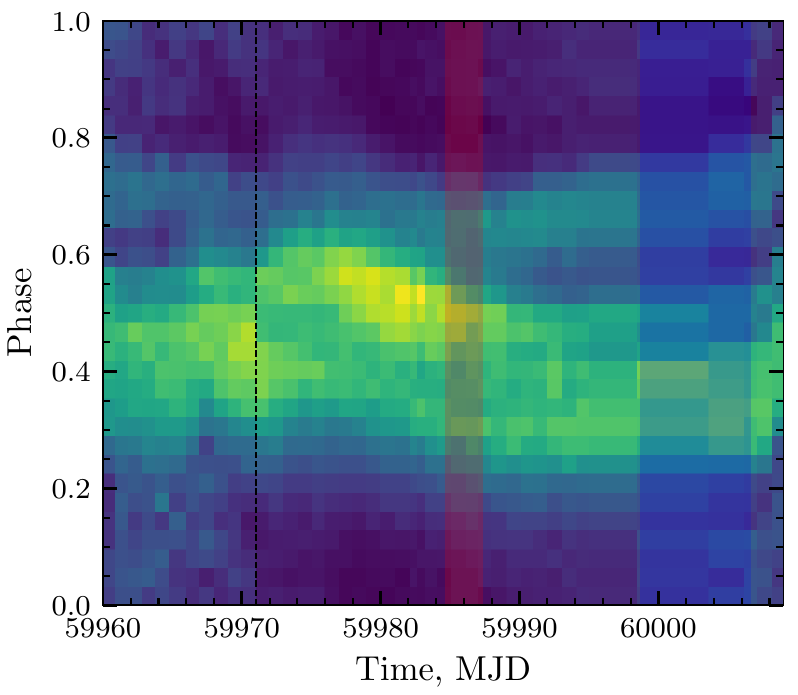}
\includegraphics[width=0.9\columnwidth]{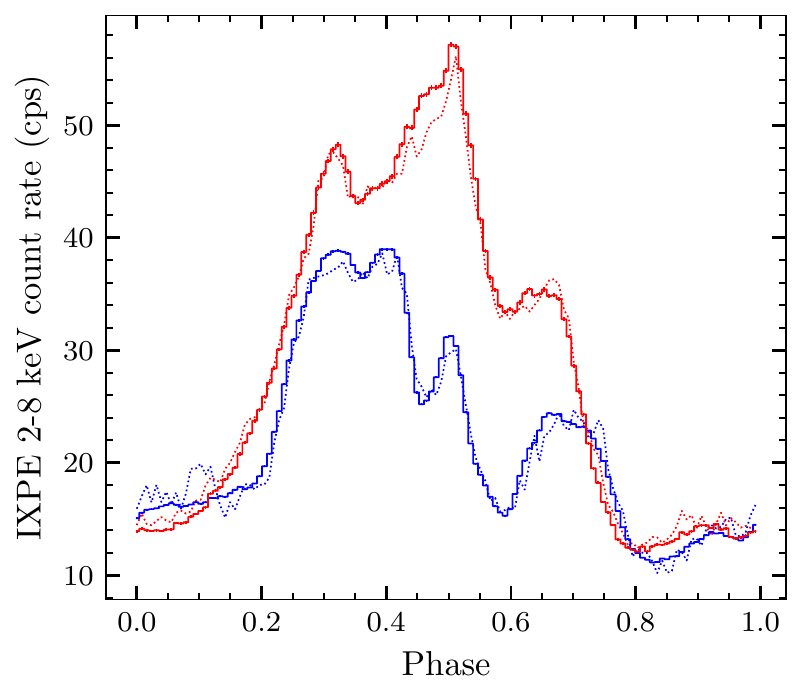}
\caption{Evolution of the pulse profiles during the outburst.  \textit{Top:}  Color-coded normalized pulse profiles as observed by \nicer. Slices in the vertical direction correspond to pulse profiles in individual \nicer observations, the shaded strips mark the times of \ixpe observations, and the vertical dashed line is the same as in Fig.~\ref{fig:lc}. 
\textit{Bottom:} Pulse profiles in the 2--8\,keV band observed by \ixpe in two observations (histogram) and \nicer during the same period (dotted line, scaled to match the \ixpe count rate).}
\label{fig:overview2}
\end{figure}

Properties of the binary remain relatively unexplored in the X-ray band as only a few relatively faint Type I outbursts typical for Be X-ray binaries have been observed up to now \citep{2010ATel.2527....1M,2012MNRAS.421.2407T,2013A&A...553A.103F}. 
Besides the outburst activity typical for BeXRBs, \lsv is also known for being one of the few systems in which accretion continues during quiescence, in this case at the X-ray luminosity $L_{\rm X}\sim (1.5-4)\times10^{34}$\,erg\,s$^{-1}$ \citep{1999MNRAS.306..100R,2012A&A...539A..82L}. 
We note that, considering the observed spin period and luminosity, the accretion in quiescence is likely powered by a cold, non-ionized disk \citep{2017A&A...608A..17T}, although wind accretion cannot be excluded. 

The transient was mainly active in 2010 and 2011 \citep{2010ATel.2527....1M,2010ATel.2537....1F,2012MNRAS.421.2407T,2013A&A...553A.103F}, reaching peak luminosities of $(1-5)\times10^{36}$\,erg\,s$^{-1}$ (here and below we adopt the revised distance estimate of 2.4\,kpc). Based on observations during this time, a tentative orbital period of $\sim150$\,d and the presence of a cyclotron line at $\sim30$\,keV, implying a magnetic field of $B\sim3\times10^{12}$\,G, have been reported by \citet{2012MNRAS.421.2407T}.  
No evidence for such a feature was, however, found by \citet{2013A&A...553A.103F} using the same data and observations of the 2011 outburst.
The reported pulse profiles were relatively simple and almost sine-like throughout the 0.3--60\,keV energy range, although some luminosity-dependent structures can be identified in the 3--15\,keV range, where counting statistics are highest \citep{2012MNRAS.421.2407T}. 
In particular, a relatively sharp dip following the main peak also reported by \citet{2012PASJ...64...79U} can be noted. 
This has been interpreted as obscuration of the emission region by the accretion stream.
The relatively low observed luminosity and a rather simple pulse-profile shape indicate that during these observations the source likely resided in the subcritical accretion regime, when the emission came directly from a hotspot and not from an accretion column expected to arise at higher luminosities \citep{1976MNRAS.175..395B}. 
On the other hand, \citet{2013A&A...553A.103F} investigated in detail the evolution of the spectral energy distribution with luminosity and conclude that the spectral curvature observed at the highest luminosities may be attributed to the transition to a radiative-pressure-dominated accretion regime and the onset of an accretion column at $L_{\rm X}\sim2\times10^{36}$\,erg\,s$^{-1}$, which they argue is consistent with theoretic expectations assuming the magnetic field strength estimated from the  observed cyclotron line energy. 
We conclude thus that there are no solid constraints on either the magnetic field or the emission-region geometry for this object.

Most recently \rx became active  in December 2022 \citep{Nakajima2022} when another Type I outburst similar to those in 2010--2011 was observed. 
The source then entered a giant outburst phase in January--February 2023, peaking at a luminosity of $\sim4\times10^{37}$\,erg\,s$^{-1}$ \citep{Pal2023,Salganik23ATel}, several times brighter than previously observed.
Extensive monitoring by several facilities, including \nustar and \nicer, allowed \citet{Coley2023} and \citet{Salganik23} to detect transitions in the spectral and timing properties of the source around MJD~59971 (January 27, 2023) and MJD~59995 (February 20, 2023). These are shown in Fig.~\ref{fig:lc}  and interpreted by \citet{Salganik23} as marking the  transition to and from a supercritical accretion regime.
Here we focus on the results obtained with the \textit{Imaging X-ray Polarimeter Explorer} \citep[\ixpe,][]{Weisskopf2022}, which observed the source in both states. 
With the launch of \ixpe, X-ray polarimetry became a new observational window to study accreting XRPs. It can be used to obtain independent constraints on their geometrical parameters through pulse-phase-resolved polarimetric analysis, which is the main objective of the current work.

The paper is organized as follows. 
In Sect. \ref{sec:obs}, we provide a summary of the observations used and briefly discuss the analysis procedures adopted. 
In Sect. \ref{sec:results} we discuss the \ixpe results in more detail and put them in the context of the results of outburst monitoring by \nicer. 
We modeled the data, provide  constraints on the pulsar's geometry, and discuss the broader implications of the  results in Sect. \ref{sec:model}.
We summarize our findings in Sect.~\ref{sec:sum}.

\section{Observations and data analysis}
\label{sec:obs}

\ixpe observed the source twice, at flux levels that differed by a factor of two. 
The first observation (Obs.~1), with ObsID 02250401, was carried out between MJD~59984.65--59987.40 (132\,ks effective exposure, with around 3.8\,M source counts in the 2-8\,keV band in total). 
The second observation (Obs.~2), with ObsID 02250501, was carried out between MJD~59998.66--60006.66 (373 ks effective exposure, with 8.3\,M source counts in total). 
We also used \nicer observations complemented with \textit{Fermi}/Gamma-ray Burst Monitor (GBM) measurements of source spin frequency \citep{2020ApJ...896...90M} to characterize the evolution of the soft pulse profile shape over the outburst in order to ensure accurate absolute pulse-phase alignment of the \ixpe data. 
Finally, we made use of the \textit{Swift}/BAT 15--50 keV light curve.\footnote{\url{https://swift.gsfc.nasa.gov/results/transients/weak/LSVp4417/}}
In the following section, we briefly describe relevant properties, analysis procedures, and the results for each instrument.

\subsection{\ixpe}

The \ixpe is a joint mission of NASA\ and the Italian Space Agency launched on December 9, 2021. It consists of three identical grazing-incidence telescope and detector modules operating in the 2--8 keV energy band. 
Each telescope comprises an X-ray mirror assembly and a polarization-sensitive detector unit equipped with a gas-pixel detector \citep{2021AJ....162..208S,2021APh...13302628B}, and provides imaging polarimetry with a time resolution better than 10~$\mu$s over the detector-limited field of view of $12\farcm9\times12\farcm9$. 
A detailed description of the observatory and its performance is given in \citet{Weisskopf2022}. The Level 2 data were processed with the {\sc ixpeobssim} package \citep{Baldini2022} version 30.2.3\footnote{\url{https://github.com/lucabaldini/ixpeobssim}} using the Calibration database released on November 17, 2022 (v12). 
Source photons were collected from a circular region with a radius $R_{\rm src}$ of 1\farcm6 centered on the source position determined by fitting a Gaussian function to the raw count map. 
The background appears to be negligible in both observations (a typical background count rate from a region of the same size in the 2--8\,keV band of \ixpe is $\sim0.02$ count\,s$^{-1}$ and the observed source count rate is $\ge10$\,count\,s$^{-1}$ at all times), and thus its contribution was ignored in the analysis \citep{Di_Marco_2023}. Taking into account the high number of source counts even in individual phase bins and the low background level, we also did not employ track weighing or acceptance corrections.

For the timing analysis, the photons' arrival times were corrected to the Solar System barycenter using the {\tt barycorr} task. 
No binary correction was done as at the time of writing the orbital parameters of the system were still not known. 
The pulsar ephemerides were then obtained for each of the \ixpe observations using the phase-connection technique \citep{1981ApJ...247.1003D} and are reported in Table~\ref{tab:eph}. 
The absolute phase alignment between the two \ixpe observations was done using the peak at around phase 0.5, which appears to be present throughout most of the outburst, as indicated by \nicer monitoring and discussed below. 
The ephemerides were then used directly to either produce pulse profiles or to generate a set of good-time-interval (GTI) files to define the phase intervals for pulse-phase-resolved analysis.
Using the \texttt{pcube} routine in \textsc{ixpeobssim} \citep{Baldini2022} in a broad 2--8\,keV band, we extracted binned Stokes parameters $I$, $Q$, and $U$ \citep{2015APh....64...40K}, taking the modulation factor of the instrument into account.
We define normalized Stokes parameters  as $q=Q/I$ and $u=U/I$. 
The polarization degree (PD) and the polarization angle (PA) can be obtained using standard formulae: PD=$\sqrt{q^2+u^2}$ and PA=$\frac{1}{2}\arctan(u/q)$. 

\begin{table}
    \centering
    \caption{Pulsar ephemerides for the two \ixpe observations. }
    \begin{tabular}{ccc}
   \hline
      \hline
    Parameter   & \ixpe1 & \ixpe2\\
    \hline
      $T_0$ (MJD)  & 59984.64718 & 59998.65768  \\
       $\nu$ (mHz)  & 4.8484(8) & 4.8670(1)  \\
       $\dot{\nu}$ (s$^{-2}$)  & $3(2)\times10^{-11}$ & $1.42(3)\times10^{-11}$  \\
       $\ddot{\nu}$ (s$^{-3}$) &   $-1(2)\times10^{-16}$ &  $-8(1)\times10^{-18}$ \\
          \hline
    \end{tabular}
\tablefoot{The uncertainties are reported at a 1$\sigma$ confidence level. The reference epoch, $T_0,$ is fixed to the arrival time of the first pulse.}
    \label{tab:eph}
\end{table}

\subsection{\nicer}

In addition to absolute phasing, monitoring of pulse profile shape changes can  also be a useful probe for possible changes in the accretion regime associated with the onset of an accretion column \citep{2013A&A...551A...1R,2018ApJ...863....9W,2020MNRAS.491.1857D}, and put the snapshot \ixpe observations in a broader context. 
All available \nicer observations were processed using the \texttt{nicerl2} task, and then light curves in the 1--10, 4--7, and 7--10\,keV energy bands were extracted using the \texttt{nicerl3-lc} script. 
The extracted light curves were then corrected to the Solar System barycenter and folded, assuming a single-phase model based on the spin-frequency measurements of the source done by \textit{Fermi}/GBM\footnote{\url{https://gammaray.nsstc.nasa.gov/gbm/science/pulsars/lightcurves/rxj0440.html}} as follows.
First, we interpolated raw frequency measurements to obtain a smooth function characterizing the frequency evolution of the source with time. 
This interpolated function was then used to calculate the absolute phase of each pulse within the period covered by the observations. 
That is, the arrival time of each subsequent pulse was calculated using the arrival time and local frequency of the previous one (for the first pulse, the phase was set arbitrarily). 
Finally, we calculated a reference epoch and folding frequency for each of the \nicer observations using the obtained interpolated functions. 
The reference phase was selected such that the narrow peak at phase $\sim0.5$ visible in both \ixpe observations occurs at the same phase as in simultaneous \nicer data. 
The result presented in Fig.~\ref{fig:overview2} exhibits no major regular phase drifts and is consistent with the correlation-based alignment procedure outlined in \citet{2020MNRAS.491.1857D}. 
We conclude therefore that, despite uncertainty in the orbital parameters of the system and rapid observed spin-up, the observed spin-frequency evolution implies that the sharp peak around phase 0.5 is indeed the same feature in both \ixpe observations, and thus their absolute phase alignment obtained above is indeed correct.

\section{Results} 
\label{sec:results}

\begin{figure} 
\centering
\includegraphics[width=0.95\columnwidth]{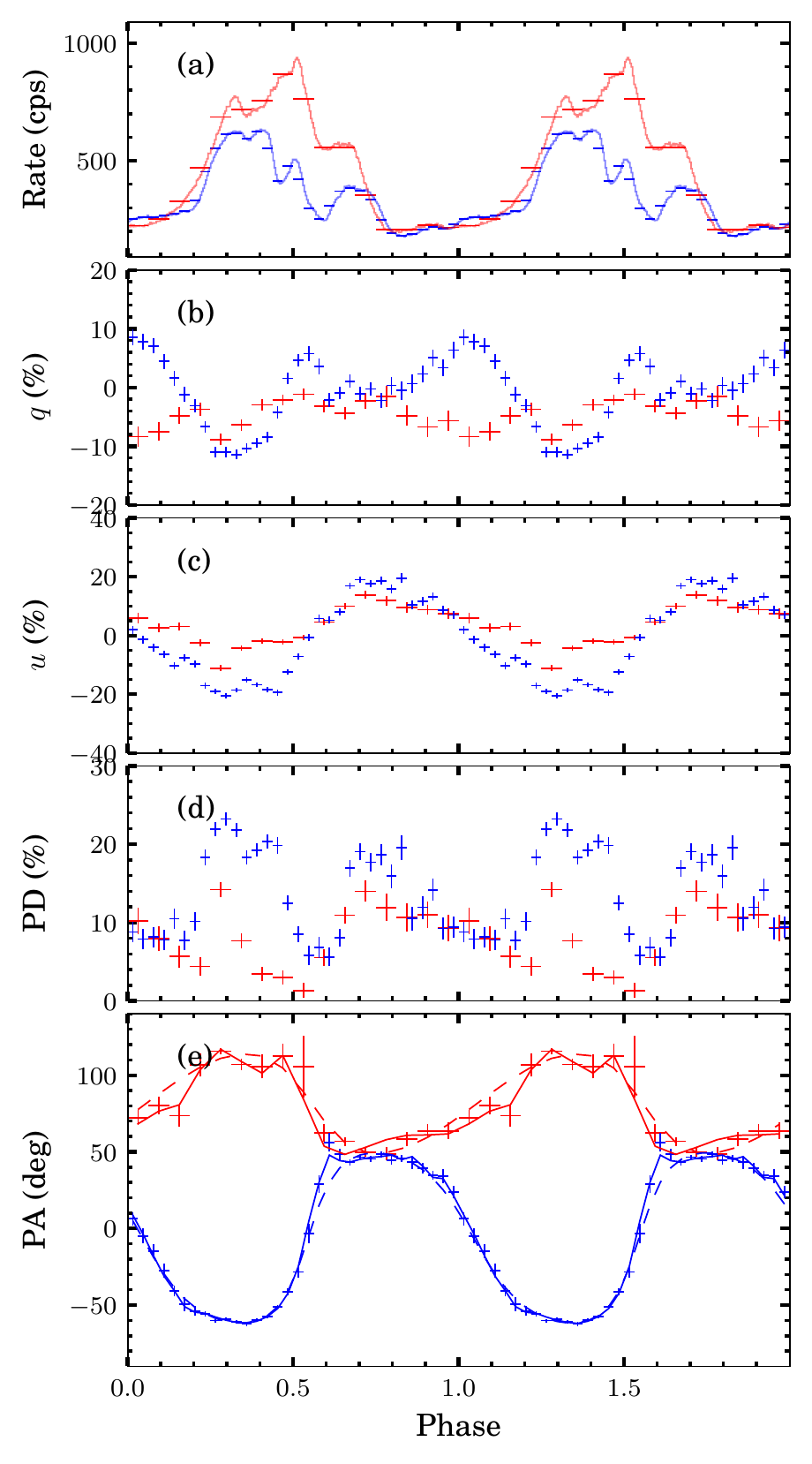}
\caption{Pulse-phase dependence of the  source flux, the normalized Stokes parameters $q=Q/I$ and $u=U/I$, the PD, and the PA for the first (red) and second (blue) \ixpe observations, respectively. 
In panel (e), the lines show the RVM best-fit model for each observation individually with no extra components (dashed, Sect.~\ref{sec:rvm}) and a joint fit including a constant polarization component (solid, Sect.~\ref{sec:alter}).}
\label{fig:pdpa}
\end{figure}

As with other accreting pulsars observed by \ixpe so far \citep{2022NatAs...6.1433D,Tsygankov23,Mushtukov23,Malacaria23,2023ApJ...947L..20F}, the average polarization in the 2--8 keV band observed from the source is low. 
We measured a PD of $4.4\%\pm0.2\%$ at PA=$79\degr\pm2\degr$ in Obs.~1 and PD=$4.9\%\pm0.2\%$, PA=$-59\degr\pm2\degr$ in Obs.~2 (uncertainties here and throughout the manuscript are reported at a  $1\sigma$ confidence level unless stated otherwise). 
More interesting is the pulse-phase dependence of the observed polarization properties. 
As a subsequent step, therefore, we conducted a phase-resolved polarimetric analysis. 
The results for both observations are presented in Fig.~\ref{fig:pdpa}. 
We also verified that spectro-polarimetric analysis with \textsc{xspec} \citep{Arn96} using a simple absorbed Comptonization model (\texttt{comptt} with parameters similar to those reported by \citet{Salganik23ATel} for the soft component) gives consistent results, only weakly affected by the assumed spectral model. 
The motivation for the choice of the binned analysis over spectro-polarimetry is discussed below in Sect.~\ref{sec:model}. 
The polarization is detected with a significance exceeding $3\sigma$ in 14 out of 16 phase bins in Obs.~1, and in all 32 phase bins of Obs.~2. 
The higher quality of the data in the latter case is related to a significantly longer exposure that allowed us to collect more photons (despite a factor of two lower flux) and a higher average PD.

As is evident from Fig.~\ref{fig:pdpa}, both the observed pulse profiles and the polarization properties appear to be drastically different between the two observations. 
The PD is significantly  larger in Obs.~2, reaching 25\%, while it remains below 15\% in Obs.~1. 
The profile of the PD (see Fig.~\ref{fig:pdpa}d) exhibits similar features, such as a peak at phase 0.3 and a secondary broad peak at phase 0.7--0.8, while the third peak of the PD at phase 0.4 present in Obs.~2 is not present in Obs.~1. 
The observed changes in PA phase dependence between the two observations are even more noteworthy. 
The PA as a function of phase not only appears to be in antiphase in the two observations, but  the amplitude of variations also changes by a factor of two. 
Another important point one could make here is the remarkable comparative simplicity of the PA phase dependence in both observations, especially in the second one (see Fig.~\ref{fig:pdpa}e). 
This can be contrasted with the complex pulse profile that exhibits multiple peaks varying with energy (see Fig.~\ref{fig:pdpa}a) and also significantly differs between the two observations (even if some common features  such as narrow peaks at phases $\sim$0.3, 0.5 and 0.7 and dips around phases 0.6 and 0.9 can be identified). 
It is also worth noting that the absolute flux around the pulse minimum remains almost constant whereas the maximum flux changes significantly. 
As a consequence, the pulsed fraction decreases from $\sim$66\% in Obs.~1 to $\sim$56\% in Obs.~2, remaining, nevertheless, unusually high \citep[see also][]{Salganik23}. 
On the other hand, the PA exhibits in both cases almost sinusoidal modulation with no extra features despite the statistics being definitively sufficient for them to be detected.

\section{Modeling}
\label{sec:model} 

The  dramatic changes observed in pulse-profile shape and polarization properties are, in fact, not totally unexpected given the likely transition to the supercritical accretion regime \citep{Salganik23} and definitively deserve more detailed analysis. 
Considering the lack of reliable model predictions for the PD, we focus below mainly on the analysis of the PA phase dependence.

\begin{figure}
\centering
\includegraphics[width=0.9\columnwidth]{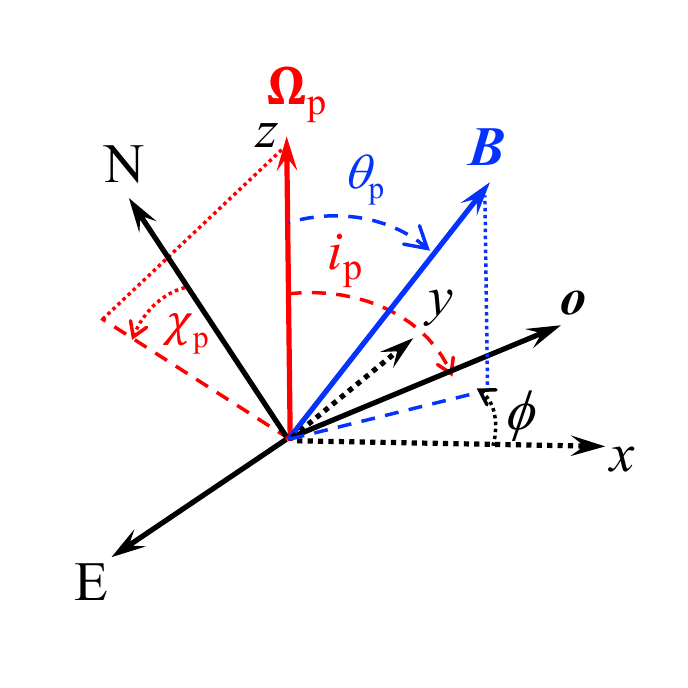}
\caption{Geometry of the pulsar and main parameters of the RVM. 
The pulsar angular momentum, $\vec{\Omega}_{\rm p}$, makes an angle $i_{\rm p}$ with respect to the line of sight (along vector $\vec{o}$). 
The angle $\theta_{\rm p}$ is between magnetic dipole, $\vec{B}$, and the NS angular momentum $\vec{\Omega}_{\rm p}$. 
Pulsar phase $\phi$ is the azimuthal angle of vector $\vec{B}$ in the plane $(x,y)$ perpendicular to  $\vec{\Omega}_{\rm p}$. 
The position angle $\chi_{\rm p}$ is the angle measured counterclockwise from the projection of $\vec{\Omega}_{\rm p}$ on the plane of the sky (N-E) and the direction to the north N. }
\label{fig:geometry}
\end{figure}

\begin{figure}
\centering
\includegraphics[width=\columnwidth]{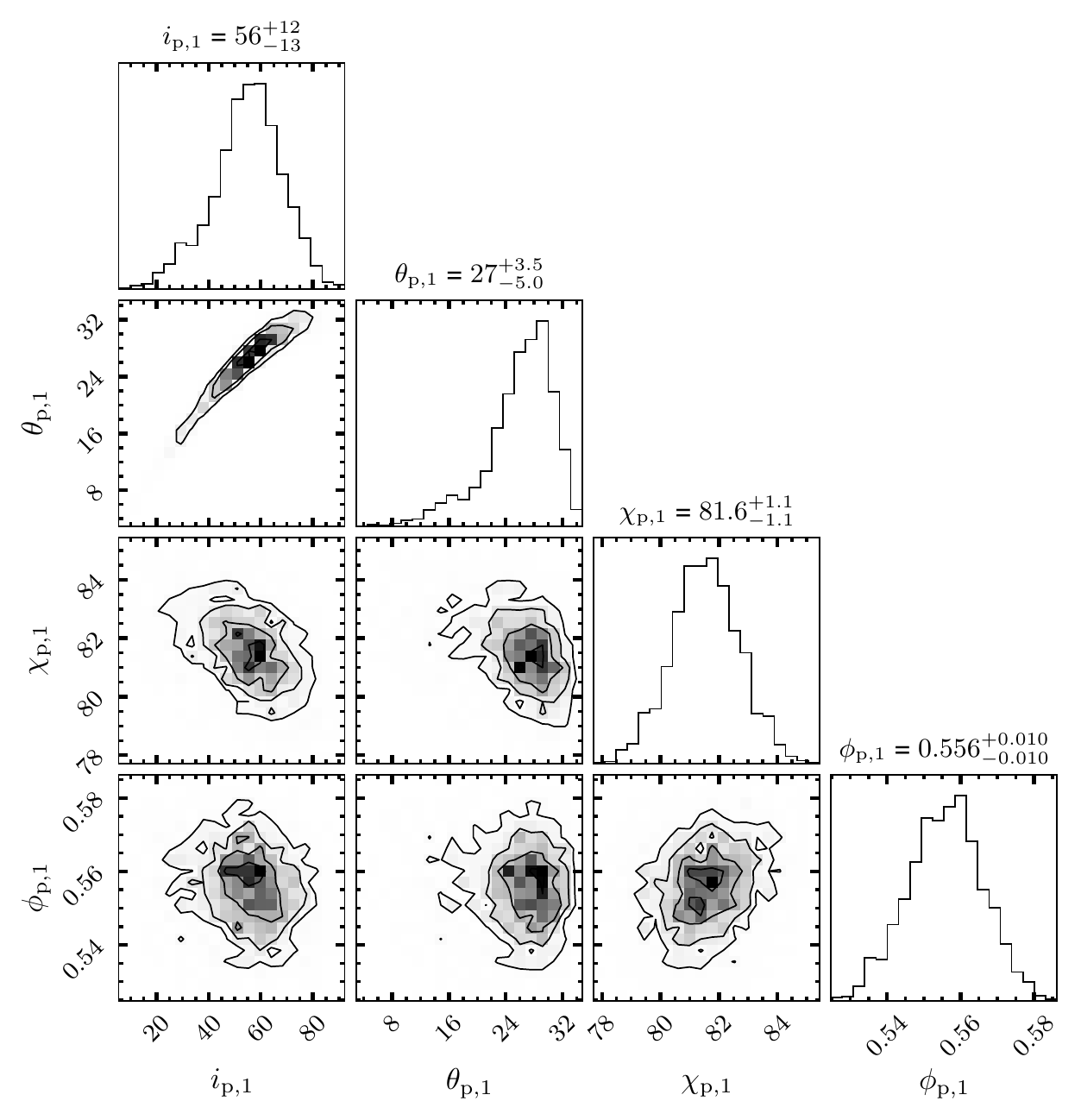}
\includegraphics[width=\columnwidth]{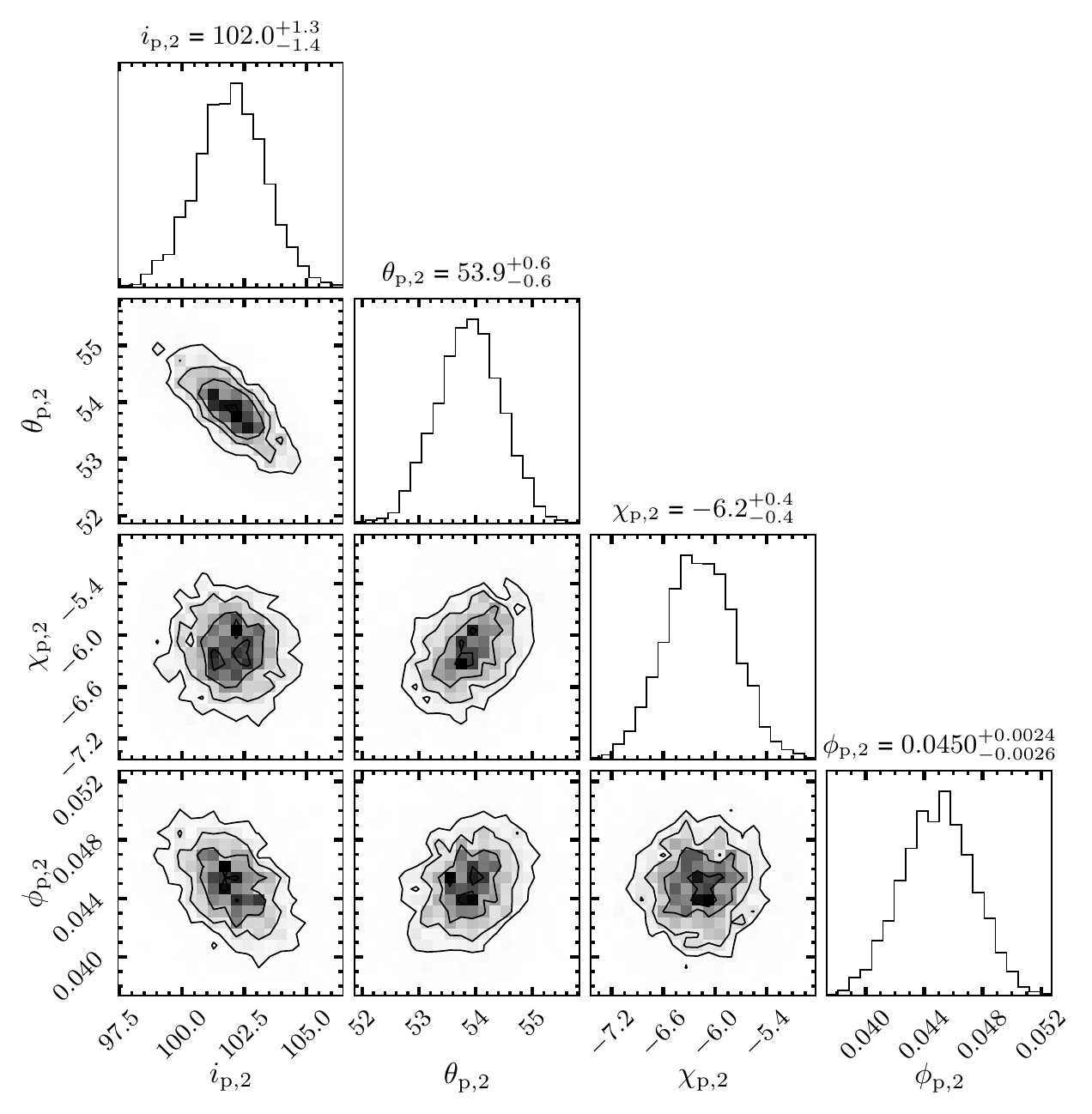}
\caption{Corner plots for the RVM fit for each observation individually without additional components. 
The results for Obs.~1 (top) and Obs.~2 (bottom) are shown. 
RVM parameters $i_{\rm p}$, $\theta_{\rm p}$, and $\chi_{\rm p}$ are in degrees.
}
\label{fig:corner_sa}
\end{figure}

\subsection{Rotating vector model}
\label{sec:rvm}

The  remarkably simple evolution of the PA with pulse phase $\phi$ observed up to now with \ixpe \citep{2022NatAs...6.1433D,Tsygankov22,Tsygankov23} is likely related to the alignment of the PA to the projection of the magnetic dipole on the sky due to vacuum polarization \citep{1978SvAL....4..117G, 1979JETP...49..741P,2000MNRAS.311..555H,2022NatAs...6.1433D,2023MNRAS.519.5902G}. 
This allowed us to constrain the basic geometry (see Fig.~\ref{fig:geometry}) of the pulsar using the rotating vector model (RVM; \citealt{1969ApL.....3..225R,Poutanen2020}):
\begin{equation} \label{eq:pa_rvm}
\tan (\chi\!-\!\chi_{\rm p})\!=\! \frac{-\sin \theta_{\rm p}\ \sin [2\pi(\phi\!-\!\phi_{\mathrm p})]}
{\sin i_{\rm p} \cos \theta_{\rm p}\!  - \! \cos i_{\rm p} \sin \theta_{\rm p}  \cos [2\pi(\phi\!-\!\phi_{\mathrm p})] } .
\end{equation} 
Here $\chi(\phi)$ is the prediction of the RVM for the PA, $\chi_{\rm p}$ is the position angle (measured counterclockwise from the direction to the north) of the pulsar angular momentum, $i_{\rm p}\in(0\degr,180\degr)$ is the inclination of the pulsar spin to the line of sight, $\theta_{\rm p}\in(0\degr,90\degr)$ is the angle between the magnetic dipole and the spin axis, and $\phi_{\mathrm p} \in (0,1)$ is the phase when the northern magnetic pole is closest to the observer.

Considering the rather different PA phase dependence in both observations, we first applied this model to each observation separately using the same Markov chain Monte Carlo (MCMC) procedure as in \citet{2022NatAs...6.1433D}, directly fitting observed PA values in individual phase bins. 
The best-fit model to the PA data is depicted  in Fig.~\ref{fig:pdpa}e. The corner plots \citep{corner} characterizing the interdependence of model parameters and showing their best-fit values are presented in Fig.~\ref{fig:corner_sa}. 
The agreement of the best-fit model with the data is striking, particularly for Obs.~2, where the data quality is the best among all pulsars observed to date by \ixpe. 
Indeed, there are only minor residuals around phase 0.6 corresponding to minimal PD values and thus having the lowest significance.

On the other hand, direct comparison of the obtained RVM parameters for the two observations implies significantly different pulsar geometry, which was not really expected. 
First, we see a rather dramatic change in the pulsar position angle, $\chi_{\rm p}$, by roughly  90\degr. 
This change might not mean that the pulsar has turned by 90\degr\ on the sky, but  could also result from the polarization-mode switch from X to O or vice versa. 
Second, there appears to be a large change in the pulsar inclination, $i_{\rm p}$, which varies from 50\degr--70\degr\ in Obs.~1 to about 100\degr\ in Obs.~2. 
The apparent change in the $\chi_p$ value is also accompanied by the change of the zero phase, $\phi_{\mathrm p}$, by half a period. 
Finally, the magnetic obliquity, $\theta_{\rm p}$, has changed from $\sim$30\degr\ to $\sim$54\degr. 
While the transition from supercritical to subcritical accretion regime \citep{Salganik23} could be expected to lead to a switch of the dominant polarization mode, it does not explain the  significant changes observed at other angles. 
We emphasize that the fact that the observed PA variations are well described by a RVM implies that the PA is defined by the magnetic field structure at the polarization radius \citep{2022NatAs...6.1433D} and thus is likely unrelated to local changes in emission-region geometry. 
On the other hand, it is, of course, difficult to imagine the orientation of the neutron star changing on such a short timescale. Moreover, time-resolved analysis similar to that described above revealed no significant geometry changes within either observation. 
We considered, therefore, an alternative explanation to the very peculiar behavior of the PA.

\subsection{Two-component polarization model}
\label{sec:alter}

Given the observed changes in the spectral hardness \citep{Salganik23} and pulse profiles, the potential presence of an additional polarized component with different properties in one or both observations could be considered as a natural explanation. 
We attempted, therefore, to single out this component using the spectro-polarimetric analysis of \ixpe data. 
Unfortunately, we found that the results are  inconclusive. This is mainly due to the fact that the available 2--8\,keV spectra do not allow one to reliably disentangle  the broad continuum components reported by \cite{Salganik23}; consequently, it it necessary to undertake broadband spectro-polarimetric analysis to get meaningful results. 
A joint analysis of \ixpe, \nicer,  \textit{SRG}/ART-XC, and \textit{Insight-HXMT} data is ongoing and will be reported elsewhere. 
However, some estimates can be obtained  by using binned \ixpe products alone, as we discuss below.

\begin{figure} 
\centering
\includegraphics[width=0.9\columnwidth]{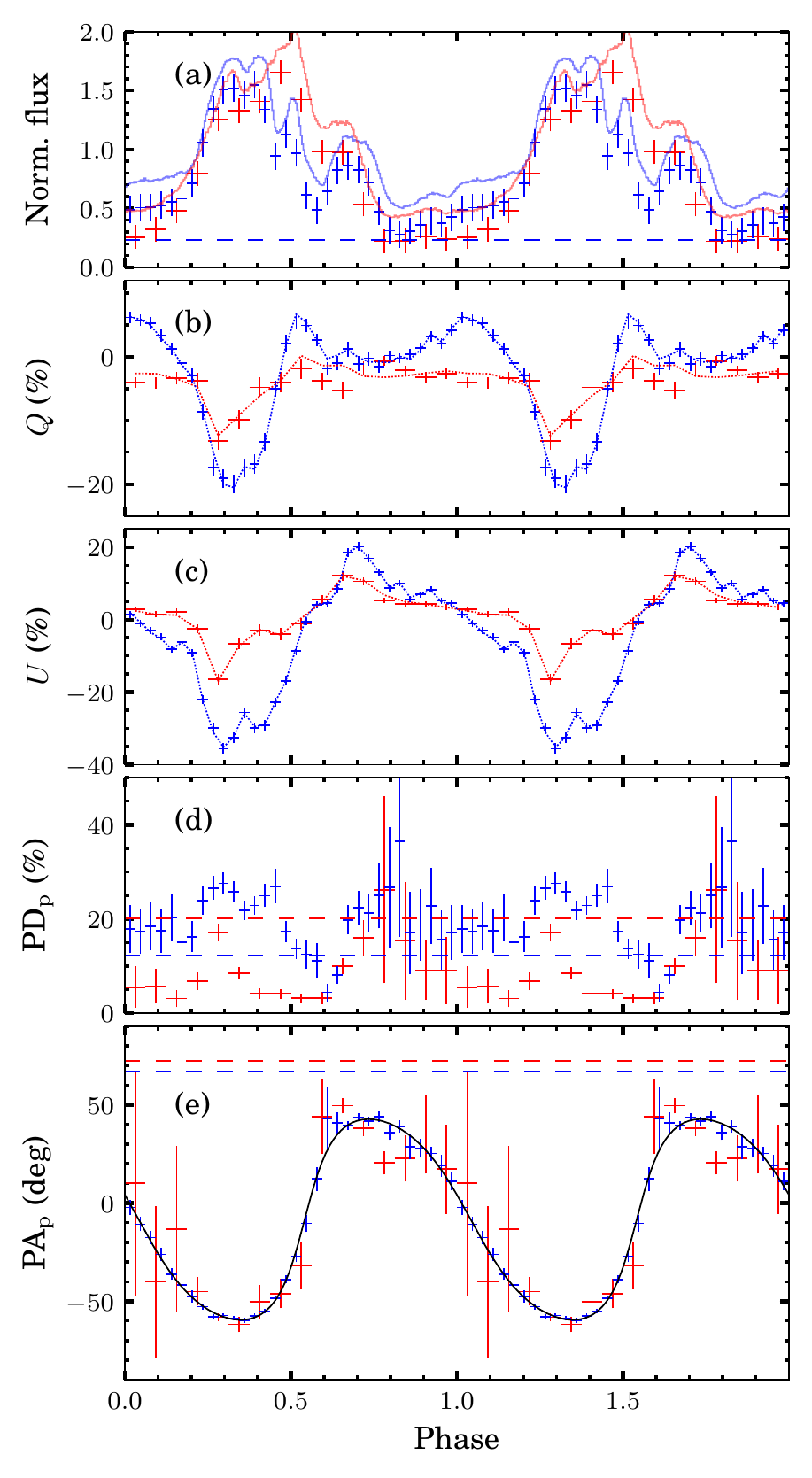}
\caption{Results for the two-component model corresponding to the best-fit parameters listed in Fig.~\ref{fig:sub_corner}.  
Panel (a) shows the total normalized flux (solid lines), the mean value for the constant component ($I_{\mathrm c,1}=I_{\mathrm c,2}=0.23$, horizontal dashed lines), and the flux of the variable component accounting for the uncertainty in the constant component (crosses). 
Panels (b) and (c) show the observed Stokes parameters normalized to the average flux (crosses) and the best-fit model (dotted lines). 
Panels (d) and (e) show the PD and PA for constant (horizontal dashed lines) and variable (crosses) components, respectively.  
The solid line in panel (e) matches the solid set of lines in Fig.~\ref{fig:pdpa}e with the constant component subtracted. 
Red and blue symbols and lines correspond to Obs.~1 and Obs.~2, respectively.}
\label{fig:sub_pdpa}
\end{figure}

Plotting Stokes parameters on the $(q,u)$ plane, we realized that there is a certain similarity between the two observations, but the amplitude of the variations is larger in Obs.~2 and the data points are shifted relative to each other. 
This supports the idea that, in addition to the polarized radiation coming from the pulsar directly, there is a component that does not depend on phase (or at least depends less on it than the variable pulsar component). 
A similar conclusion could be reached if one considers the ``off-pulse'' as background in both observations (i.e., resulting in a more similar phase dependence for both the PD and PA). 
We attempted, therefore, to disentangle the two components through modeling of the observed Stokes parameters. 
This can be done by expressing the absolute Stokes parameters for each observation as a sum of the variable component described by the RVM and an additional constant component: 
\begin{eqnarray}  
\label{eq:two_comp}
I(\phi) &=& I_{\mathrm c} + I_{\mathrm p}(\phi) , \nonumber \\
Q(\phi) &=& Q_{\mathrm c} + P_{\mathrm p}(\phi)I_{\mathrm p}(\phi)\cos[2\chi(\phi)] , \\
U(\phi) &=& U_{\mathrm c} + P_{\mathrm p}(\phi)I_{\mathrm p}(\phi) \sin[2\chi(\phi)]  .  \nonumber
\end{eqnarray} 
By $I$, $Q,$ and $U,$ we can assume here that the observed Stokes parameters are normalized to the average flux value with indices denoting the constant (c) and pulsed (p) components, $P_{\mathrm p}$ is the PD of the variable component, and its PA $\chi$ is given by Eq.~\eqref{eq:pa_rvm}.
The Stokes parameters $(Q_{\mathrm c},U_{\mathrm c})$ are related to the PD, $P_{\mathrm c}$, and the flux, $I_{\mathrm c}$, of the constant component, 
\begin{equation}  \label{eq:qu_consta} 
 Q_{\mathrm c}= P_{\mathrm c} I_{\mathrm c} \cos(2\chi_{\mathrm c}),\quad U_{\mathrm c}= P_{\mathrm c} I_{\mathrm c} \sin(2\chi_{\mathrm c}) ,
\end{equation} 
with its PA being $\chi_{\mathrm c}=(1/2)\arctan(U_{\mathrm c}/Q_{\mathrm c})$. 
The polarized flux of the variable component was computed as 
\begin{equation}  
\label{eq:pf1} 
P_{\mathrm p}I_{\mathrm p}(\phi) = \sqrt{[Q(\phi)-Q_{\mathrm c}]^2+[U(\phi)-U_{\mathrm c})]^2 } . 
\end{equation} 
The expected PD and PA of the total emission could then be obtained from the summed Stokes parameters of both components and could be compared with the observed values. 
The null hypothesis is now that the geometry of the pulsar does not change between the observations and the observed changes in the polarization properties are related to the presence of an additional, unpulsed, polarized component. 
This means that a single RVM could fit the variable $\mbox{PA}(\phi)$ for both observations.  

\begin{figure*} 
\centering
\includegraphics[width=\textwidth]{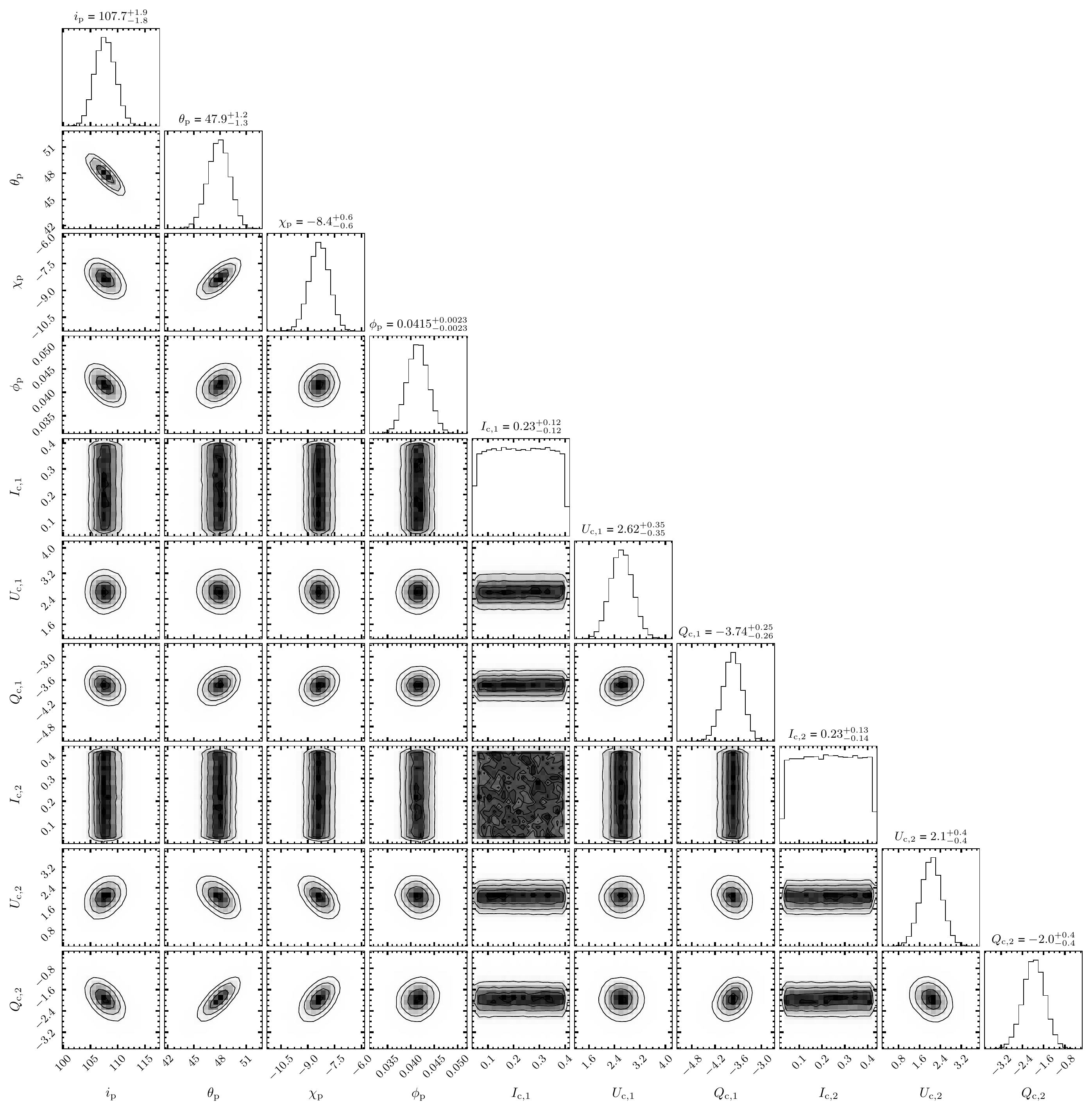}
\caption{Corner plots for the simultaneous fit of both observations including the unpulsed polarization component contribution (parameters correspond to a set of solid lines in Fig.~\ref{fig:pdpa}e and lines in Fig.~\ref{fig:sub_pdpa}). 
The RVM parameters $i_{\rm p}$, $\theta_{\rm p}$, and $\chi_{\rm p}$ are in degrees. 
The Stokes parameters $Q_{\mathrm c,i}$ and $U_{\mathrm c,i}$ are expressed in percent of the average flux, while $I_{\mathrm c,i}$ are fractions of the average flux.
The uncertainties are reported at a $1\sigma$ confidence level. }
\label{fig:sub_corner}
\end{figure*}

In practice, this can be done by including six additional parameters $(I_{\mathrm c,i},Q_{\mathrm c,i},U_{\mathrm c,i})$ corresponding to the Stokes parameters (normalized to the average flux) of the constant component in two observations ($i=1,2$) in the model so that the right part of Eq.~\eqref{eq:two_comp} is fully defined. 
The four RVM parameters and $(I_{\mathrm c,i},Q_{\mathrm c,i},U_{\mathrm c,i})$ can then be estimated by comparing the model prediction with the observed $Q,U$ values. In this case modeling is done directly in $Q, U$ space, so observed PA or PD values are not used directly.
To obtain model parameters and their uncertainties, we used MCMC sampling as implemented in the \textsc{emcee} package \citep{2013PASP..125..306F}, assuming uniform priors for all parameters except for $i_{\rm p}$ and $\theta_{\rm p}$, where flat priors for cosine of the angles were assumed. 
The likelihood was calculated using $\chi^2$ statistics for $Q,U$ (as the uncertainties of the observed Stokes parameters are normally distributed), and set to negative infinity for parameters outside of ranges defined above for the RVM and for $P_{\mathrm c}>1$ and $P_{\mathrm p}(\phi)>1$ to account for prior knowledge of their possible values. The results are presented in   Figs.~\ref{fig:sub_pdpa} and \ref{fig:sub_corner}.

We emphasize that the main role in the analysis above is played by the Stokes parameters $Q_{\mathrm c},U_{\mathrm c}$ (normalized to the averaged flux) of the constant component. 
The obtained RVM parameters $i_{\rm p}= 108\degr\pm2\degr$, $\theta_{\rm p}=48\degr\pm1\degr$, $\chi_{\rm p}=-8\fdg4\pm0\fdg6$, and $\phi_{\mathrm p}=0.041\pm0.002$ do not depend at all on the assumption about $I_\mathrm{c}$ and they are well constrained within the framework of the adopted two-component polarization model. 
From the best-fit Stokes parameters of the constant component  of $Q_{\mathrm c,1}=-3.7\%\pm0.3\%$, $U_{\mathrm c,1}=2.6\%\pm0.3\%$ and  $Q_{\mathrm c,2}=-2.0\%\pm0.4\%$, $U_{\mathrm c,2}=2.1\%\pm0.4\%$, we can get its PAs,  $\chi_{\mathrm c,1}=72\degr\pm2\degr$ and $\chi_{\mathrm c,2}=67\degr\pm4\degr$, and the polarized fluxes, $P_{\mathrm c,i}I_{\mathrm c,i}= \sqrt{Q_{\mathrm c,i}^2+U_{\mathrm c,i}^2}$ of $4.5\%\pm0.3\%$ and $2.9\%\pm0.4\%$, for the two observations, respectively. 
The data allowed us to constrain the polarized flux of the constant component, 
whereas the flux and the PD separately cannot be well determined. 
The limits on $I_{\mathrm c,i}$ only appear because the PDs of both components, $P_{\mathrm c}$ and $P_{\mathrm p}$, need to be less than 100\%. 
This condition translates to the limits on the flux of the constant component, $I_{\mathrm c}\ge P_{\mathrm c} I_{\mathrm c}$ and $I_{\mathrm c} \le \min[I(\phi)-P_{\mathrm p}(\phi)I_{\mathrm p}(\phi)]$, as is apparent in Fig.~\ref{fig:sub_corner}. 
The resulting limits are $I_{\mathrm c,1}\in[0.038,0.42]$ for Obs.~1 and $I_{\mathrm c,2}\in[0.013,0.44]$ for Obs.~2.
For the maximum possible $I_{\mathrm c}$, we get the minimum possible $P_{\mathrm c,1}\approx$12\% and $P_{\mathrm c,2}\approx7$\%, and $P_{\mathrm c}$ grows inversely proportional to $I_{\mathrm c}$. 
For example, for the mean $I_{\mathrm c}=0.23$, we get $P_{\mathrm c,1}\approx20\%$ and $P_{\mathrm c,2}\approx12.5\%$.  

These estimates are relevant for discussion of the physical origin of the constant component. 
Indeed, to remain independent of the pulse phase, the constant component must originate far from the neutron star. 
Possible sites include reprocessing in the matter piled up at the magnetosphere, reflection from the accretion disk or in the disk winds, scattering in the circumbinary disk of the Be star, or reprocessing in the atmosphere of the donor star itself. 
The fraction of reprocessed light is, however, on the order of 20\%. This is relatively high, so the origin of the component must be able to explain it, which favors locations relatively close to the pulsar, such as the magnetosphere and/or the accretion disk.
Considering the growing evidence for the presence of outflows launched from inner regions of the accretion disk of BeXRBs during both Type I and Type II outbursts  \citep{2019ApJ...885...18J,2019MNRAS.487.4355V,2022MNRAS.516.4844V,2022MNRAS.509.2532C,2022MNRAS.516.4844V} and an expected non-negligible effect on pulse profiles and spectra at higher accretion rates \citep{2023MNRAS.518.5457M}, scattering in a highly ionized equatorial disk wind seems to be a plausible scenario.
It has also been suggested that reprocessing in the inner disk regions and disk wind is responsible for the soft excess observed in many transient XRPs \citep{2004ApJ...614..881H}, including \rx \citep{Salganik23}, for which the soft excess was also found to be more prominent at higher luminosities. 
Larger deviations from the RVM and the higher PD of the constant component in Obs.~1, where the accretion rate and likely the outflow rate were higher, is also in line with this hypothesis. 
For an idealized case in which the scattering material lies in a plane, the polarization of the scattered component depends on the inclination to the plane normal, $i$, as PD=$\sin^2 i / (3-\cos ^2 i)$  \citep{ST85}, reaching  33\% edge-on, and is still larger than 30\% for $i>66\degr$.  
For a wind occupying a larger solid angle, the PD drops, but even for the half-opening of the wind (measured from the orbital plane) of 30\degr\ (i.e., occupying half of the sky as seen from the pulsar), the PD drops  by just a factor of 0.77, being above 22\% for $i>66\degr$.  
These estimates of the PD are comparable to the data. 
The contribution of the scattered emission to the total flux in the soft band can also be appreciable, reaching at least 10\%\ at higher accretion rates \citep{2019ApJ...885...18J}.
More accurate constraints on the fraction of scattered light and polarization of the scattered component can also be obtained through broadband spectral analysis and detailed modeling of scattering and fluorescent lines in the vicinity of the pulsar. 

We note also that a high estimated value of pulsar inclination suggests that the accretion disk is likely being viewed close to edge-on, which is expected to yield the largest polarization. 
We note that in this scenario the PA of the constant component is expected to be aligned with the position angle of the normal to the accretion disk and the orbital plane, and thus is not expected to change with the orbital phase, which is consistent with observations. 
The orientation of the accretion disk relative to the decretion disk of the Be star can be tested through optical polarimetric observations. 
Preliminary analysis of data obtained as part of our optical polarimetric campaign with the DIPol-2 high-precision polarimeter \citep{Piirola14} at the T60 telescope at Haleakala Observatory  yields a PA of the intrinsic optical polarization of 55\degr--71\degr (Nitindala A.~P. et al., in prep.). 
Such a close agreement with the X-ray PA  does not look like a coincidence and probably implies that the pulsar orbit lies close to the decretion-disk plane. 

It is interesting to discuss the phase dependence of the variable component. First, one can note that $\phi_{\mathrm p}=0$ corresponds to the northern pole coming close to the observer, while at $\phi_{\mathrm p}=0.5$ the southern pole is close. For the estimated $i_{\rm p}= 108\degr$ and $\theta_{\rm p}=48\degr$, the minimum angles between the normal to the spots and the line of sight are $i_{\rm p}-\theta_{\rm p}=60\degr$ and  $180\degr-\theta_{\rm p}-i_{\rm p}=24\degr$, respectively. 
Thus, it is not surprising that the flux has a maximum when the southern spot is closer to the observer in both observations.

\section{Summary and conclusions} 
\label{sec:sum}

We have presented the first results of \ixpe observations of the 2023 Type II outburst of the bright Be transient, \rx. 
The observations were carried out at two luminosity levels and likely captured the object in two accretion regimes associated with the presence and  absence of an accretion column \citep{1976MNRAS.175..395B,Salganik23}.
This presented a unique opportunity to probe the changes in emission-region geometry and radiative processes related to the onset of an accretion column by means of pulse-phase-resolved X-ray polarimetry.

Our analysis indeed revealed that the source is strongly polarized in both observations, with the PD exceeding 20\% in some phase bins. 
The observed PA phase dependence remains remarkably consistent with the predictions from the RVM model \citep{1969ApL.....3..225R,Poutanen2020}, but the derived geometrical parameters appear to be completely different for the two observations.  
While the observed 90\degr\ jump in the pulsar spin position angle could be attributed to a change in the dominant polarization mode associated with the transition, there is no obvious explanation for changes in other angles that amount to tens of degrees on a timescale of just a few days. 
It is difficult to imagine that such changes are associated with a true change in the orientation of the neutron star's spin and magnetic axes with respect to the observer. 
That led us to consider alternative explanations, in particular the potential presence of an additional polarized component. 

We find that the observed PA phase dependence in both observations can indeed be described with no changes in pulsar geometry if a strongly polarized (PD$\sim$10--30\%, PA$\approx70\degr$) unpulsed component is present. 
Subtracting the Stokes parameters of the constant component from the observed ones, we have derived constraints on the pulsar geometry, obtaining a pulsar inclination of $i_{\rm p}\approx108\degr$, a magnetic obliquity of  $\theta_{\rm p}\approx48\degr$, and a position angle of the pulsar spin of $\chi_{\rm p}\approx-8\fdg4$. 
Properties of the unpulsed component (e.g., its contribution to the total flux, the PD, and the PA) appear to be roughly constant between the two observations.  
We suggest that this component can be produced by scattering the pulsar radiation in a highly ionized disk wind. 
The observed PA phase dependence is consistent, and for Obs.~2 shows almost perfect agreement, with the simple RVM model used up to now to study the geometry of XRPs. 
Yet, despite formal agreement, a more detailed analysis of the two observations carried out at different epochs reveals that the situation is far more complicated. It cannot be excluded that similar complications might be relevant for studies of other XRPs observed by \ixpe. 
We conclude, therefore, that the importance of high-quality, multi-epoch polarimetric  observations, preferably accompanied by broadband spectroscopic and optical polarimetric observations, should not be underestimated.

\begin{acknowledgements}
The Imaging X-ray Polarimetry Explorer (\ixpe) is a joint US and Italian mission.  
The US contribution is supported by the National Aeronautics and Space Administration (NASA) and led and managed by its Marshall Space Flight Center (MSFC), with industry partner Ball Aerospace (contract NNM15AA18C).  
The Italian contribution is supported by the Italian Space Agency (Agenzia Spaziale Italiana, ASI) through contract ASI-OHBI-2017-12-I.0, agreements ASI-INAF-2017-12-H0 and ASI-INFN-2017.13-H0, and its Space Science Data Center (SSDC) with agreements ASI-INAF-2022-14-HH.0 and ASI-INFN 2021-43-HH.0, and by the Istituto Nazionale di Astrofisica (INAF) and the Istituto Nazionale di Fisica Nucleare (INFN) in Italy.
This research used data products provided by the \ixpe Team (MSFC, SSDC, INAF, and INFN) and distributed with additional software tools by the High-Energy Astrophysics Science Archive Research Center (HEASARC), at NASA Goddard Space Flight Center (GSFC). 
We acknowledge support from the German Academic Exchange Service (DAAD) travel grant 57525212 (VD, VFS), 
the Academy of Finland grants 333112, 349144, 349373, and 349906 (JP, SST, SVF), 
the Natural Sciences and Engineering Research Council of Canada (NSERC) and the Canadian Space Agency (JH), 
the V\"ais\"al\"a Foundation (SST), 
UKRI Stephen Hawking fellowship (AAM), 
the German Research Foundation (DFG) grant \mbox{WE 1312/59-1} (VFS),   
the Vilho, Yrj\"o and Kalle V\"ais\"al\"a Foundation, and Suomen Kulttuurirahasto (VK). 
\end{acknowledgements}

\bibliographystyle{aa}
\bibliography{bibtex.bib}

\end{document}